\documentstyle[a4,12pt]{article}
\begin{document}

\setcounter{page}{0}

\begin{flushright}
{CERN-TH.7194/94}\\
{PVAMU-HEP-94-2}
\end{flushright}

\begin{center}
{\Large\bf On Measuring $CP$ Violation in Neutral $B$-meson Decays
at the $\Upsilon (4S)$ Resonance}
\end{center}

\vspace{0.2cm}

\begin{center}
{\bf Harald FRITZSCH}\footnote{Supported in part by DFG contract No. F412/12-2
and C.E.C. project SCI-CT-91-0729}\\
{\sl Theory Division, CERN, CH-1211 Geneva 23, Switzerland}\\
{\sl and}\\
{\sl Sektion Physik der Universit${\sl\ddot a}$t M${\sl\ddot u}$nchen,
D-80333 Munich, Germany}
\end{center}

\begin{center}
{\bf Dan-di WU}\footnote{Supported in part by the U.S. Department of Energy
under contract DE-FG05-92ER40728}\\
{\sl HEP, Box 355, Prairie View A$\&$M University, Prairie View, TX 77446, USA}
\end{center}

\begin{center}
{\bf Zhi-zhong XING}\footnote{Alexander-von-Humboldt Research Fellow}\\
{\sl Sektion Physik der Universit${\sl\ddot a}$t M${\sl\ddot u}$nchen,
D-80333 Munich, Germany}
\end{center}

\vspace{0.6cm}

\begin{center}
{\bf Abstract}
\end{center}

\small

	Within the standard model we carry out an analysis of $CP$-violating
observables in neutral $B$-meson decays at the $\Upsilon (4S)$
resonance. Both time-dependent and time-integrated
$CP$ asymmetries are calculated, without special approximations,
to meet various possible measurements at symmetric and asymmetric
$e^{+}e^{-}$ $B$ factories.
We show two ways to distinguish between direct
and indirect $CP$-violating effects in the $CP$-eigenstate channels such
as $B^{0}_{d}/\bar{B}^{0}_{d}\rightarrow \pi^{+}\pi^{-}$ and
$\pi^{0} K_{S}$. Reliable knowledge of
the Cabibbo-Kobayashi-Maskawa phase and angles can in principle be
extracted from measurements of some non-$CP$-eigenstate channels,
e.g. $B^{0}_{d}/\bar{B}^{0}_{d}\rightarrow D^{\pm}\pi^{\mp}$ and
$\stackrel{(-)}{D}$$^{(*)0}K_{S}$, even in the presence of significant
final-state interactions. \\

\newpage

\normalsize

\begin{flushleft}
{\Large\bf 1. Introduction}
\end{flushleft}

	Observing $CP$ violation in the $B$-meson system and confronting
it with the predictions of the standard model is a challenging task
in particle physics. On the experimental side,
a sample of at least $10^{8}$ $B$ mesons is required
before meaningful measurements of $CP$ asymmetries can be
carried out. On the theoretical side, there are two
obstacles to reliable numerical predictions for
$CP$ asymmetries in exclusive non-leptonic $B$ decays. First, the present
knowledge of some of the underlying
electroweak parameters such as the Cabibbo-Kobayashi-Maskawa (CKM) matrix
elements [1] is insufficient. Second, intrinsic uncertainties due
to the impact of strong interactions exist.
While the former awaits more accurate measurements, the latter needs a
deeper understanding of the dynamics of the non-leptonic weak decays.

	It is well known that large $CP$ asymmetries may arise in many
exclusive decay channels of neutral $B$ mesons, either from $B^{0}$-$\bar
{B}^{0}$ mixing or via final-state strong interactions, or by a
combination of the two [2]. However, many of the previous quantitative
predictions are problematic because they have ignored or injudiciously
simplified final-state interactions [3]. In the long run, a great improvement
of these
calculations is possible to yield reliable results.
It is more instructive at present to explore
various available measurements of $CP$ asymmetries at $B$ factories, in order
to
test the unitarity of the CKM matrix and study the impact of final-state
interactions
in neutral $B$ decays in a model-independent way.

	In this letter we present a non-trivial analysis of
$CP$-violating observables in correlated decays of $B^{0}_{d}\bar{B}^{0}_{d}$
pairs at the $\Upsilon (4S)$ resonance, a clean basis of the future
symmetric and asymmetric $e^{+}e^{-}$ $B$ factories [4]. Two categories of
interesting
transitions are focused on: (1) $B^{0}_{d}$ and $\bar{B}^{0}_{d}$ decays
to $CP$ eigenstates, in which significant QCD-loop-induced (penguin)
contributions
may exist; and (2) $B^{0}_{d}$ and $\bar{B}^{0}_{d}$ decays to common
non-$CP$ eigenstates, whose amplitudes depend only on a single weak phase.
To meet various possible measurements proposed for a new generation of
$e^{+}e^{-}$ colliders running at the $\Upsilon (4S)$ [4,5], we calculate
both the time-dependent and time-integrated decay probabilities and
$CP$ asymmetries by taking final-state interactions into account. Some useful
relations between the observables and the weak and strong transition phases
are obtained.
We show two ways to distinguish between direct and
indirect $CP$-violating effects in the $CP$-eigenstate decays such as
$B^{0}_{d}/\bar{B}^{0}_{d}\rightarrow \pi^{+}\pi^{-}$ and $\pi^{0} K_{S}$.
In principle, reliable knowledge of the CKM phase and angles
can be extracted from measurements of some non-$CP$-eigenstate
decays, e.g. $B^{0}_{d}/\bar{B}^{0}_{d}\rightarrow D^{\pm}\pi^{\mp}$ and
$\stackrel{(-)}{D}$$^{(*)0}K_{S}$, even in the presence of significant
final-state strong interactions.
We also point out that
measurements of a few pure penguin modes such as
$B^{0}_{d}/\bar{B}^{0}_{d}\rightarrow \phi K_{S}$ should serve as a good
test of the existing calculations for the strong penguin diagrams. \\

\begin{flushleft}
{\Large\bf 2. Decay probabilities of $B^{0}_{d}\bar{B}^{0}_{d}$
pairs at the $\Upsilon (4S)$}
\end{flushleft}

	The unique experimental advantages of studying $b$-quark physics
at the $\Upsilon (4S)$ are well known. For either symmetric or asymmetric
$e^{+}e^{-}$ collisions, the detectors required to measure
$CP$ violation in correlated decays of $B^{0}_{d}\bar{B}^{0}_{d}$
events are sophisticated, but within the limits of the present
technology [4,5].
On the $\Upsilon (4S)$ resonance, the $B$'s are produced in a two-body
($B^{+}_{u}B^{-}_{u}$ or $B^{0}_{d}\bar{B}^{0}_{d}$) state with definite
charge parity $C=-$. The two neutral $B$ mesons mix coherently
until one of them decays. Thus one can use the semileptonic decays of
one meson to tag the flavour of the other meson decaying to a
flavour-non-specific hadronic final state.
At a centre-of-mass beam
energy above $m^{~}_{B}+m^{~}_{B^{*}}$ but below $2m^{~}_{B^{*}}$, the
$e^{+}e^{-}$
annihilation can produce $B\bar{B}^{*}$ or $B^{*}\bar{B}$ pairs, which
dominate the $b\bar{b}$ final states [5]. The $B^{*}$ ($\bar{B}^{*}$) will
decay
radiatively to the $B$ ($\bar{B}$), leaving $B\bar{B}\gamma$ with the
$B\bar{B}$ in a $C=+$ state. At a $B$ factory, both the
$B^{0}_{d}\bar{B}^{0}_{d}$ decays in the $C=-$ and $C=+$ states
are worth studying in order to search for large $CP$-violating effects.

	Supposing one neutral $B$ meson decaying to a
semileptonic state $|l^{\pm}X^{\mp}\rangle$ at (proper) time $t_{1}$ and
the other to a non-leptonic state $|f\rangle$ at time $t_{2}$, the
time-dependent
probabilities for such a joint decay can be given by [2]
\begin{equation}
\begin{array}{ccl}
{\rm Pr}(l^{\pm}X^{\mp}, t_{1}; f, t_{2})_{C} & \propto &
|A_{l}|^{2} |A_{f}|^{2} e^{-\Gamma (t_{1}+t_{2})}
\left [ \displaystyle\frac{1+|\xi_{f}|^{2}}{2} \right . \\
&  & \left . \mp \displaystyle\frac{1-|\xi_{f}|^{2}}
{2} \cos [\Delta m (t_{2}+Ct_{1})] \pm {\rm Im}\xi_{f}
\sin [\Delta m (t_{2}+Ct_{1})] \right ] \; ,
\end{array}
\end{equation}
where $C$ ($=\pm$) is the charge parity of the $B^{0}_{d}\bar{B}^{0}_{d}$
pair. Here we have defined $\Gamma\equiv (\Gamma_{1}+\Gamma_{2})/2$ and
$\Delta m\equiv m_{2}-m_{1}$, where $\Gamma_{1,2}$ and $m_{1,2}$
are the widths and masses of the $B_{d}$ mass eigenstates, $B_{1,2}$.
In obtaining Eq. (1), two good approximations
$\Delta \Gamma \equiv \Gamma_{1}-\Gamma_{2} << \Gamma$ and $\Delta \Gamma <<
\Delta m$
have been used [6]. In addition,
\begin{equation}
\begin{array}{ccl}
A_{l} & \equiv & \langle l^{+}X^{-}|H|B^{0}_{d}\rangle \; \stackrel{CPT}{==} \;
\langle l^{-}X^{+}|H|\bar{B}^{0}_{d}\rangle \; ,  \\
A_{f} & \equiv & \langle f|H|B^{0}_{d}\rangle \; , \;\;\;\; \bar{A}_{f} \;
\equiv \;
\langle f|H|\bar{B}^{0}_{d}\rangle \; , \;\;\;\; \xi_{f} \; \equiv \;
e^{-2i\phi_{B}}
\displaystyle \frac{\bar{A}_{f}}{A_{f}} \; ,
\end{array}
\end{equation}
where $\phi_{B}\equiv \arg (V_{tb}V^{*}_{td})$ is the phase
of $B^{0}_{d}$-$\bar{B}^{0}_{d}$ mixing [2,6].
For the case that one neutral $B$ meson decays to $|l^{\mp}X^{\pm}\rangle$ at
time
$t_{1}$ and the other decays
to $|\bar{f}\rangle$ (the $CP$-conjugate state of $|f\rangle$) at time $t_{2}$,
the
corresponding decay probabilities
${\rm Pr}(l^{\mp}X^{\pm},t_{1};\bar{f},t_{2})_{C}$
can be obtained from Eq. (1) by the replacements $A_{f}\rightarrow
\bar{A}_{\bar{f}}$
and $\xi_{f}\rightarrow \bar{\xi}_{\bar{f}}$ , where
\begin{equation}
\bar{A}_{\bar{f}}\; \equiv \; \langle \bar{f}|H|\bar{B}^{0}_{d}\rangle \; ,
\;\;\;\; A_{\bar{f}}\; \equiv \; \langle \bar{f}|H|B^{0}_{d}\rangle \; ,
\;\;\;\;
\bar{\xi}_{\bar{f}} \; \equiv \; e^{2i\phi_{B}}\frac{A_{\bar{f}}}
{\bar{A}_{\bar{f}}} \; .
\end{equation}

	The difference between the decay probabilities associated with
$B^{0}_{d}\rightarrow f$ and $\bar{B}^{0}_{d}\rightarrow \bar{f}$ is a
basic signal for $CP$ violation. In practice, one has to
consider the possibility of an $e^{+}e^{-}$ collider to measure the
time development of the decay
rates and $CP$ asymmetries. For a symmetric collider running at the
$\Upsilon (4S)$ resonance, the mean decay length of $B$'s is
insufficient for the measurement of
$(t_{2}-t_{1})$ [4]. On the other hand, the quantity $(t_{2}+t_{1})$ cannot
be measured in a symmetric or asymmetric storage ring operating at the
$\Upsilon (4S)$, unless the bunch lengths are much shorter than the decay
lengths [4,5]. Therefore, only the time-integrated measurements are available
at
a symmetric $B$ factory. Integrating ${\rm Pr}(l^{\pm}X^{\mp},t_{1};
f,t_{2})_{C}$ over $t_{1}$ and $t_{2}$, we obtain
\begin{equation}
{\rm Pr}(l^{\pm}X^{\mp},f)_{-} \; \propto \; |A_{l}|^{2}|A_{f}|^{2}\left [
\frac{1+|\xi_{f}|^{2}}{2}\mp \frac{1}{1+x^{2}_{d}}
\frac{1-|\xi_{f}|^{2}}{2} \right ] \; ,
\end{equation}
and
\begin{equation}
{\rm Pr}(l^{\pm}X^{\mp},f)_{+} \; \propto \; |A_{l}|^{2}|A_{f}|
^{2} \left [\frac{1+|\xi_{f}|^{2}}{2}\mp \frac{1-x^{2}_{d}}
{(1+x^{2}_{d})^{2}}\frac{1-|\xi_{f}|^{2}}{2}
\pm\frac{2x_{d}}{(1+x^{2}_{d})^{2}}{\rm Im}\xi_{f} \right ] \; ,
\end{equation}
where $x_{d}\equiv \Delta m/\Gamma \sim 0.7$ is a measurable of
$B^{0}_{d}$-$\bar{B}^{0}_{d}$ mixing [7].
For an asymmetric collider running in this energy region, one might want to
integrate Eq. (1) only over $(t_{2}+t_{1})$ in order to measure the
development of decay probabilities with $\Delta t\equiv (t_{2}-t_{1})$ [4,5].
In this case, we obtain
\begin{equation}
{\rm Pr}(l^{\pm}X^{\mp},f;\Delta t)_{-} \; \propto \; |A_{l}|^{2}|A_{f}|^{2}
e^{-\Gamma |\Delta t|}\left [\frac{1+|\xi_{f}|^{2}}{2}\mp
\frac{1-|\xi_{f}|^{2}}{2}\cos (\Delta m \Delta t)
\pm {\rm Im}\xi_{f}\sin (\Delta m \Delta t) \right ] \; ,
\end{equation}
and
\begin{eqnarray}
{\rm Pr}(l^{\pm}X^{\mp},f;\Delta t)_{+} & \propto & |A_{l}|^{2}|A_{f}|^{2}
e^{-\Gamma |\Delta t|} \left [ \displaystyle\frac{1+|\xi_{f}|^{2}}{2}
\mp\displaystyle\frac{1}
{\sqrt{1+x^{2}_{d}}}\displaystyle\frac{1-|\xi_{f}|^{2}}{2}\cos (\Delta m
|\Delta t| +
\phi_{x_{d}}) \right . \nonumber \\
&  & \left . \pm \displaystyle\frac{1}{\sqrt{1+x^{2}_{d}}}{\rm Im}\xi_{f}\sin
(\Delta m |\Delta t| +\phi_{x_{d}})\right ] \; ,
\end{eqnarray}
where $\phi_{x_{d}}\equiv \tan^{-1}x_{d} \sim 35^{\circ}$.

	From Eqs. (4)-(7) one can straightforwardly obtain the decay
probabilities associated with $B^{0}_{d}\rightarrow \bar{f}$ and
$\bar{B}^{0}_{d}\rightarrow \bar{f}$.
Corresponding to the possible measurements for joint
$B^{0}_{d}\bar{B}^{0}_{d}$ decays at symmetric (S) and
asymmetric (A) $B$ factories, we define the $CP$-violating
asymmetries as
\begin{equation}
{\cal A}^{\rm S}_{C} \; \equiv \; \frac{{\rm Pr} (l^{-}X^{+},f)_{C}
-{\rm Pr} (l^{+}X^{-},\bar{f})_{C}}
{{\rm Pr} (l^{-}X^{+},f)_{C}+{\rm Pr} (l^{+}X^{-},\bar{f})_{C}} \; ,
\end{equation}
and
\begin{equation}
{\cal A}^{\rm A}_{C}(\Delta t) \; \equiv \; \frac{{\rm Pr}(l^{-}X^{+},f;\Delta
t)
_{C}-{\rm Pr}(l^{+}X^{-},\bar{f};\Delta t)_{C}}
{{\rm Pr}(l^{-}X^{+}, f;\Delta t)_{C}+{\rm Pr}(l^{+}X^{-}, \bar{f};\Delta t)
_{C}} \; .
\end{equation}
In the following, we shall calculate ${\cal A}^{\rm S}_{C}$ and
${\cal A}^{\rm A}_{C}(\Delta t)$ for two categories of interesting neutral $B$
decays and explore relations between the observables and the weak
and strong transition phases in them. \\

\begin{flushleft}
{\Large\bf 3. $CP$ asymmetries in $B_{d}$ decays to $CP$ eigenstates}
\end{flushleft}

	We first consider the $B^{0}_{d}$ and $\bar{B}^{0}_{d}$ decays
to $CP$ eigenstates (i.e. $|\bar{f}\rangle =n^{~}_{CP}|f\rangle$ with
$n^{~}_{CP}=\pm 1$),
such as $J/\psi K_{S}, \pi^{+}\pi^{-}$, and $\pi^{0}K_{S}$.
With the phase convention $CP|B^{0}_{d}\rangle =|\bar{B}^{0}_{d}\rangle$, we
have $A_{\bar{f}}=n^{~}_{CP}A_{f}, \bar{A}_{\bar{f}}=n^{~}_{CP}\bar{A}_{f}$,
and $\bar{\xi}_{\bar{f}}=1/\xi_{f}$.
For symmetric and asymmetric $e^{+}e^{-}$ collisions
at the $\Upsilon (4S)$, the
corresponding $CP$ asymmetries in $(B^{0}_{d}\bar{B}^{0}_{d})_{C}
\rightarrow (l^{\pm}X^{\mp})f$ are given by
\begin{equation}
\begin{array}{ccl}
{\cal A}^{\rm S}_{-} & = & \displaystyle\frac{1}{1+x^{2}_{d}}{\cal U}_{f} \; ,
\\
{\cal A}^{\rm S}_{+} & = & \displaystyle\frac{1-x^{2}_{d}}{(1+x^{2}_{d})^{2}}
{\cal U}_{f}+ \displaystyle\frac{2x_{d}}{(1+x^{2}_{d})^{2}}
{\cal V}_{f} \; ;
\end{array}
\end{equation}
and
\begin{equation}
\begin{array}{ccl}
{\cal A}^{\rm A}_{-}(\Delta t) & = & {\cal U}_{f}
\cos (\Delta m \Delta t) +{\cal V}_{f}
\sin (\Delta m \Delta t) \; ,  \\
{\cal A}^{\rm A}_{+}(\Delta t) & = & \displaystyle\frac{1}{\sqrt{1+x^{2}_{d}}}
\left [{\cal U}_{f} \cos (\Delta m |\Delta t| + \phi_{x_{d}})
+{\cal V}_{f} \sin (\Delta m |\Delta t| + \phi_{x_{d}}) \right ] \; ,
\end{array}
\end{equation}
where
\begin{equation}
{\cal U}_{f} \; =\; \frac{1-|\xi_{f}|^{2}}{1+|\xi_{f}|^{2}} \; , \;\;\;\;\;
{\cal V}_{f} \; =\; \frac{-2{\rm Im}\xi_{f}}{1+|\xi_{f}|^{2}} \; .
\end{equation}
Non-vanishing ${\cal U}_{f}$ and ${\cal V}_{f}$ imply the $CP$ asymmetry in the
decay amplitude (i.e. $|\xi_{f}|\neq 1$) and the one from interference between
decay and mixing, respectively. From the above equations one can observe a few
interesting features.

	(1) ${\cal A}^{\rm S}_{-}$ is a
pure measure of direct $CP$ violation, while ${\cal A}^{\rm S}_{+}$
contains both direct and indirect $CP$ asymmetries.
A combination of the measurements for ${\cal A}^{\rm S}_{\pm}$ can in principle
distinguish between direct and indirect $CP$-violating effects in neutral
$B$-meson
decays [8]. From Eqs. (10) and (12), we obtain
\begin{equation}
\begin{array}{ccl}
{\cal U}_{f} & = & (1+x^{2}_{d}) {\cal A}^{\rm S}_{-} \; , \\
{\cal V}_{f} & = & \displaystyle\frac{(1+x^{2}_{d})^{2}}{2x_{d}}\left [{\cal
A}^{\rm S}_{+}
- \displaystyle\frac{1-x^{2}_{d}}{1+x^{2}_{d}}{\cal A}^{\rm S}_{-}\right ] \; .
\end{array}
\end{equation}

	(2)  Compared with
the time-integrated $CP$ asymmetry in incoherent $B^{0}_{d}$ and
$\bar{B}^{0}_{d}$ decays to $CP$ eigenstates [9] (e.g.
in a hadronic production environment or in high energy $e^{+}e^{-}$
reactions [10]):
\begin{equation}
{\cal A} \; =\; \frac{1}{1+x^{2}_{d}}{\cal U}_{f} + \frac{x_{d}}
{1+x^{2}_{d}}{\cal V}_{f} \; ,
\end{equation}
the direct and
indirect parts of ${\cal A}^{\rm S}_{+}$ have the additional dilution factors
$(1-x^{2}_{d})/(1+x^{2}_{d})\sim 0.34$ and $2/(1+x^{2}_{d})\sim 1.34$,
respectively.

	(3) Both ${\cal A}^{\rm A}_{\pm}(\Delta t)$ are composed of
direct and indirect $CP$ violation and have the following relation:
\begin{equation}
{\cal A}^{\rm A}_{+}(\Delta t)\; =\; \frac{1}{\sqrt{1+x^{2}_{d}}}
{\cal A}^{\rm A}_{-}\left (|\Delta t| +\frac{\phi_{x_{d}}}{\Delta m}\right ) \;
{}.
\end{equation}
In contrast with
${\cal A}^{\rm A}_{-}(\Delta t)$,
the asymmetry ${\cal A}^{\rm A}_{+}(\Delta t)$ has a dilution factor
in its magnitude
$1/\sqrt{1+x^{2}_{d}}\sim 0.82$, and a positive shift in its phase
$\phi_{x_{d}}\sim 35^{\circ}$.

	(4) If $|\xi_{f}|=1$, only
the asymmetries via mixing remain in ${\cal A}^{\rm A}_{\pm}(\Delta t)$. As a
signal of the existence of direct $CP$ violation,
the deviation of $|\xi_{f}|$ from unity can also be probed
by measuring the time development of the asymmetries.
In particular, one can extract the information on direct $CP$ violation
with the help of
\begin{equation}
{\cal A}_{-}^{\rm A}\left (\frac{n\pi}{\Delta m}\right )\; =\;
(-1)^{n}{\cal U}_{f} \; ,
\end{equation}
where $n=0,\pm 1,\pm 2$, and so on.

	From the measurements of ${\cal A}^{\rm S}_{\pm}$ or
${\cal A}^{\rm A}_{\pm}(\Delta t)$ one can obtain the information
on $|A_{f}|, |\xi_{f}|$, and ${\rm Im}\xi_{f}$ for the decays
$B^{0}_{d}\rightarrow f$ and $\bar{B}^{0}_{d}\rightarrow n^{~}_{CP}f$.
At the quark level, most of the
neutral $B$ decays to $CP$ eigenstates occur through the transitions
$b\rightarrow (Q\bar{Q})q$ (with $Q=u,c,t$ and $q=d,s$) and their
flavour-conjugate processes. With the help of
the CKM unitarity $V_{ub}V^{*}_{uq}+V_{cb}V^{*}_{cq}+V_{tb}V^{*}_{tq}=0$,
the decay amplitudes $A_{f}$ and $\bar{A}_{\bar{f}}$ can
be symbolically expressed as
\begin{equation}
\begin{array}{ccl}
A_{f} & = & \left [A_{u}e^{i(-\phi_{u}+\delta_{u})}+A_{c}e^{i(-\phi_{c}
+\delta_{c})}\right ] e^{-i\phi_{K}}\; ,  \\
\bar{A}_{\bar{f}} & = & n^{~}_{CP}\left [A_{u}e^{i(\phi_{u}+\delta_{u})}
+A_{c}e^{i(\phi_{c}+\delta_{c})}\right ] e^{i\phi_{K}} \; ,
\end{array}
\end{equation}
where $\phi_{u}\equiv \arg(V_{ub}V^{*}_{uq})$ and $\phi_{c}\equiv
\arg(V_{cb}V^{*}_{cq})$ are the $b$-decay (weak) phases, $\delta_{u,c}$ are
the corresponding strong phases, $A_{u,c}$ are the full (real)
amplitudes calculated to first order in the weak interactions and (in
principle)
to all orders in the strong interactions,
and $\phi_{K}$ is a weak phase associated with the possible $K^{0}$-$\bar
{K}^{0}$ mixing in the final state
($\phi_{K}=\arg (V_{cs}V^{*}_{cd})$ when $|f\rangle$ contains
a single $K_{S}$ or $K_{L}$, and $\phi_{K}=0$ when $|f\rangle$ is of
zero strangeness).
Defining $h\equiv A_{c}/A_{u}$, $\delta \equiv \delta_{c}-\delta_{u}$,
and $\phi_{M}\equiv \phi_{B}-\phi_{K}$, one obtains
\begin{eqnarray}
|A_{f}|^{2} & = & A^{2}_{u}\left [1+2h\cos (\delta +\phi_{u}-
\phi_{c}) +h^{2}\right ] \; , \nonumber \\
|\xi_{f}|^{2} & = & \frac{1+2h\cos (\delta -\phi_{u}+\phi_{c})+h^{2}}
{1+2h\cos (\delta +\phi_{u}-\phi_{c}) +h^{2}} \; , \\
{\rm Im}\xi_{f} & = & n^{~}_{CP}\frac{\sin 2(\phi_{u}-\phi_{M})
+2h\cos \delta \sin (\phi_{u}+\phi_{c}-2\phi_{M})
+h^{2}\sin 2(\phi_{c}-\phi_{M})}
{1+2h\cos (\delta +\phi_{u}-\phi_{c})+h^{2}} \; . \nonumber
\end{eqnarray}
In Eq. (18), three relations are given between the observables ($|A_{f}|$,
$|\xi_{f}|$, and ${\rm Im}\xi_{f}$) and the weak and strong transition
parameters ($\phi_{u,c}$, $\phi_{M}$,
$A_{u,c}$, and $\delta$). If the relevant weak phases
have been well determined elsewhere, one may use these relations to probe
$A_{u,c}$ and $\delta$,
in order to obtain the information on final-state interactions.
In principle, the
decay amplitudes $A_{f}$ and $\bar{A}_{\bar{f}}$ can be evaluated with the
help of effective weak Hamiltonians [11] and QCD [12].
The experimental
determination of $A_{u,c}$ and $\delta$ will provide a test of
the theoretical calculations.

	There are two categories of interesting
decay modes, in which no significant entanglement exists between the
tree-level and penguin contributions:

	(1) For the decay modes with dominant tree-level amplitudes such as
$B^{0}_{d}/\bar{B}^{0}_{d}\rightarrow J/\psi K_{S}$ and $D^{+}D^{-}$,
one can safely neglect the component $A_{u}$ in $A_{f}$ and
$\bar{A}_{\bar{f}}$. As a result,
Eq. (18) is simplified as
\begin{equation}
|A_{f}|=A_{c}\; , \;\;\;\;\; |\xi_{f}|=1\; , \;\;\;\;\; {\rm Im}
\xi_{f}=n^{~}_{CP}\sin 2(\phi_{c}-\phi_{M}) \; ,
\end{equation}
where only the $CP$
violation from mixing remains. Taking $B^{0}_{d}$ versus
$\bar{B}^{0}_{d}\rightarrow
J/\psi K_{S}$ for example, we show the relative size between ${\cal A}^{\rm
A}_{-}
(\Delta t)$ and ${\cal A}^{\rm A}_{+}(\Delta t)$ as well as
between ${\cal A}^{\rm S}
_{-}$ and ${\cal A}^{\rm S}_{+}$ in Fig. 1. Obviously one of the angles of the
CKM unitarity triangle, $\beta \equiv \arg
(-V^{*}_{cb}V_{cd}/V^{*}_{tb}V_{td})$,
can be reliably determined from the measurements of ${\cal A}^{\rm
A}_{\pm}(\Delta t)$
or ${\cal A}^{\rm S}_{+}$ with the help of the
relation\footnote{Note that here $n^{~}_{CP}=-1$ since $J/\psi K_{S}$ is a
$CP$-odd state.}
${\rm Im}\xi_{J/\psi K_{S}}=\sin (2\beta)$.

	(2) For the pure penguin transitions such as
$B^{0}_{d}/\bar{B}^{0}_{d}\rightarrow \phi K_{S}$ and $K^{0}\bar{K}^{0}$,
$A_{u,c}$ contain no tree-level components and may be more easily calculated.
Using perturbative QCD and simplifying final-state hadronization,
the quantities $A_{u,c}$ and $\delta_{u,c}$ have been estimated
for some charmless exclusive decay modes [13]. Those rough results
can give one a feeling of ballpark numbers to be expected.
Since the electromagnetic penguin transitions such as
$B^{0}_{d}\rightarrow \gamma K^{*0}$ have been observed recently [14],
a further study of the pure strong penguin decays would be very useful.
In such decays a comparison between the theoretical and experimental results
of $h$ and $\delta$ will provide a good test of the understanding of the
strong penguin diagrams. In practice,
$B^{0}_{d}$ versus $\bar{B}^{0}_{d}\rightarrow \phi K_{S}$ might be most
promising, since their branching ratios are on the order of $10^{-5}$ [15], a
level at which current $B$ experiments start to observe rare decays
[16]. \\

\begin{flushleft}
{\Large\bf 4. $CP$ asymmetries in $B_{d}$ decays to non-$CP$ eigenstates}
\end{flushleft}

	Now we consider the case that $B^{0}_{d}$ and $\bar{B}^{0}_{d}$
decay to a common non-$CP$ eigenstate (i.e. $|\bar{f}\rangle\neq n^{~}_{CP}
|f\rangle$), but
their amplitudes $A_{f}$ ($A_{\bar{f}}$) and $\bar{A}_{\bar{f}}$
($\bar{A}_{f}$) contain only a single weak phase.
Most of such decays occur through the quark transitions
$\stackrel{(-)}{b}\rightarrow u\bar{c}\stackrel{(-)}{q}$ and
$c\bar{u}\stackrel{(-)}{q}$ (with $q=d,s$), and a typical
example is $B^{0}_{d}/\bar{B}^{0}_{d}\rightarrow
\stackrel{(-)}{D}$$^{(*)0}K_{S}$ as illustrated in Fig. 2.
In this case, no measurable direct
$CP$ violation arises in the decay amplitudes since $|\bar{A}_{\bar{f}}|
=|A_{f}|, |\bar{A}_{f}|=|A_{\bar{f}}|$, and $|\bar{\xi}_{\bar{f}}|=
|\xi_{f}|$ (see Eq. (23)). For symmetric and asymmetric $e^{+}e^{-}$
collisions at the $\Upsilon (4S)$ resonance, the corresponding
$CP$ asymmetries in such decay modes are given by
\begin{equation}
\begin{array}{ccl}
{\cal A}^{\rm S}_{-} & = & 0 \; , \\
{\cal A}^{\rm S}_{+} & = & \displaystyle\frac{-2x_{d}{\rm
Im}(\xi_{f}-\bar{\xi}_{\bar{f}})}
{2+x^{2}_{d}+x^{4}_{d}+x^{2}_{d}(3+x^{2}_{d})|\xi_{f}|^{2}-2x_{d}{\rm Im}
(\xi_{f}+\bar{\xi}_{\bar{f}})} \; ;
\end{array}
\end{equation}
and
\begin{equation}
\begin{array}{ccl}
{\cal A}^{\rm A}_{-}(\Delta t) & = & \displaystyle\frac{-{\rm Im}(\xi_{f}
-\bar{\xi}_{\bar{f}})\sin(\Delta m\Delta t)}
{(1+|\xi_{f}|^{2})+F(\xi_{f}, \bar{\xi}_{\bar{f}}, \Delta m\Delta t)} \; , \\
{\cal A}^{\rm A}_{+}(\Delta t) & = & \displaystyle\frac{-{\rm Im}(\xi_{f}
-\bar{\xi}_{\bar{f}})\sin (\Delta m |\Delta t| + \phi_{x_{d}})}
{\sqrt{1+x^{2}_{d}}(1+|\xi_{f}|^{2})+F(\xi_{f}, \bar{\xi}_{\bar{f}},
\Delta m |\Delta t| +\phi_{x_{d}})} \; ,
\end{array}
\end{equation}
where $F$ is a function defined as
\begin{equation}
F(x,y,z)\; \equiv \; (1-|x|^{2})\cos z -{\rm Im}(x+y) \sin z \; .
\end{equation}
Compared with the $CP$ asymmetries in neutral $B$ decays to $CP$
eigenstates, here ${\cal A}^{\rm S}_{+}$ is a pure measure of $CP$ violation
via $B^{0}_{d}$-$\bar{B}^{0}_{d}$ mixing. Note that the evolution of
${\cal A}^{\rm A}_{\pm}(\Delta t)$ slightly deviates from the
harmonic oscillation. From the above equations we see that the quantities
$|\xi_{f}|$,
${\rm Im}\xi_{f}$, and ${\rm Im}\bar{\xi}_{\bar{f}}$ can be determined if the
measurements of ${\cal A}^{\rm S}_{+}$ or ${\cal A}^{\rm A}_{\pm}(\Delta t)$
are carried out at future $B$ factories.
It should be noted that in some previous studies, $\bar{\xi}_{\bar{f}}
=\xi_{f}^{*}$ was taken in order to simplify final-state interactions and allow
numerical estimates. Certainly this is a very special condition
and only valid for a few decay modes. The processes shown in Fig. 2 provide an
example where $\bar{\xi}_{\bar{f}}\neq \xi^{*}_{f}$, since the final states
$\stackrel{(-)}{D}$$^{(*)0}K_{S}$ contain both $I=0$ and $I=1$ isospin
configurations.

	Symbolically the decay amplitudes $A_{f}$ ($A_{\bar{f}}$) and
$\bar{A}_{\bar{f}}$ ($\bar{A}_{f}$) can be written as
\begin{equation}
\begin{array}{ccl}
A_{f} & = & A_{\delta} e^{i(-\phi +\delta)} e^{-i\phi_{K}} \; , \;\;\;\;
\bar{A}_{\bar{f}} \; = \; A_{\delta} e^{i(\phi +\delta)} e^{i\phi_{K}} \; ; \\
A_{\bar{f}} & = & A_{\tilde{\delta}} e^{i(-\tilde{\phi}+\tilde{\delta})}
e^{-i\phi_{K}}\; , \;\;\;\; \bar{A}_{f} \; = \; A_{\tilde{\delta}}
e^{i(\tilde{\phi}+\tilde{\delta})} e^{i\phi_{K}} \; ,
\end{array}
\end{equation}
where $\phi\equiv \arg (V_{cb}V^{*}_{uq})$ and $\tilde{\phi}\equiv
\arg (V_{ub}V^{*}_{cq})$ (with $q=d,s$) are the $b$-decay (weak) phases,
$\delta$
and $\tilde{\delta}$ are the corresponding strong phases, $A_{\delta}$
and $A_{\tilde{\delta}}$ are the real (positive) hadronic amplitudes,
and $\phi_{K}$ is the possible $K^{0}$-$\bar{K}^{0}$ mixing phase in the
final states as defined in Eq. (17). With the notation
$\Delta \delta \equiv \tilde{\delta}-\delta$, we obtain
\begin{equation}
\begin{array}{ccl}
{\rm Im}\xi_{f} & = & \displaystyle\frac{A_{\tilde{\delta}}}{A_{\delta}}
\sin (\Delta \delta +\phi +\tilde{\phi} -2\phi_{M}) \; , \\
{\rm Im}\bar{\xi}_{\bar{f}} & = &
\displaystyle\frac{A_{\tilde{\delta}}}{A_{\delta}}
\sin (\Delta \delta - \phi -\tilde{\phi}+2\phi_{M}) \; .
\end{array}
\end{equation}
Or equivalently,
\begin{equation}
\begin{array}{ccl}
{\rm Im}(\xi_{f}+\bar{\xi}_{\bar{f}}) & = & 2|\xi_{f}|\sin \Delta \delta
\cos (\phi +\tilde{\phi}-2\phi_{M}) \; , \\
{\rm Im}(\xi_{f}-\bar{\xi}_{\bar{f}}) & = & 2|\xi_{f}|\cos \Delta \delta
\sin (\phi +\tilde{\phi}-2\phi_{M}) \; .
\end{array}
\end{equation}

	From these relations one can reliably determine $\Delta \delta$ and
$(\phi +\tilde{\phi}-2\phi_{M})$, once $|\xi_{f}|$, ${\rm Im}\xi_{f}$,
and ${\rm Im}\bar{\xi}_{\bar{f}}$ have been measured in experiments.
This is really interesting because we do not need to ignore the presence
of significant final-state interactions in these decays. In Ref. [17],
$B^{0}_{d}\rightarrow \stackrel{(-)}{D}$$^{0(*)}K_{S}$ and their $CP$-conjugate
processes have been considered to probe the angles of the CKM
unitarity triangle $\alpha$ and $\beta$ with the help of an approximate
form of Eq. (24). Certainly one can also apply Eq. (24) or (25) to
some other related decay modes such as
$f=D^{(*)\pm}\pi^{\mp}, \stackrel{(-)}{D}$$^{(*)0}\pi^{0},F^{(*)\pm}K^{\mp}$,
and $\stackrel{(-)}{D}$$^{(*)0}J/\psi$. On the experimental side, to
detect such charmed channels should be a little easier than to detect
those without charm. \\

\begin{flushleft}
{\Large\bf 5. Discussion and conclusion}
\end{flushleft}

	To meet various possible measurements at symmetric and asymmetric
$e^{+}e^{-}$ $B$ factories, we have analysed both time-dependent and
time-integrated $CP$-violating asymmetries in correlated decays of
$B^{0}_{d}$ and $\bar{B}^{0}_{d}$ mesons
at the $\Upsilon (4S)$ resonance. A parallel discussion can be given for
joint $B^{0}_{s}\bar{B}^{0}_{s}$ decays at the $\Upsilon (5S)$. Because of the
very large $B^{0}_{s}$-$\bar{B}^{0}_{s}$ mixing predicted by the standard
model ($x_{s}\sim O(10)$ [18]), the observable size of
time-integrated $CP$ asymmetries at the $\Upsilon (5S)$
will be considerably diluted.
On the other side, the time-dependent measurements are also difficult
for $B^{0}_{s}$ and $\bar{B}^{0}_{s}$ decays due to the expected
rapid rate of oscillation.
In fact, it is almost impossible to
accumulate sufficient $B_{s}$ events at the $\Upsilon (5S)$ for
the study of $CP$-violating effects. Using hadron collisions
or high energy $e^{+}e^{-}$ reactions (e.g. at the $Z$ peak)
for producing beauty mesons, one might be able to measure the proper time
evolution of some $B_{s}$ decays in the future. Then Eqs. (18) and
(24) remain useful for analysing the weak and strong transition
phases in them. As an independent test of the CKM picture of
$CP$ violation, the study of incoherent $B_{s}$
decays such as $B^{0}_{s}/\bar{B}^{0}_{s}\rightarrow J/\psi \phi$
and $\phi\stackrel{(-)}{D}$$^{0}$ is quite helpful.

	It is worth while at this point to remark that in Eq. (17)
(or Eq. (23)) the decay amplitudes $A_{f}$ and
$\bar{A}_{\bar{f}}$ are parametrized in terms of their weak phases,
where all the strong interaction contributions
are included in $A_{u,c}$ and $\delta_{u,c}$ (or
$A_{\delta}, A_{\tilde{\delta}}, \delta$ and $\tilde{\delta}$).
Reliably evaluating these strong-interaction quantities is
a serious theoretical challenge.
In the literature most of the numerical
calculations are done with the help of effective weak Hamiltonians and
factorization approximations, where all the long-distance
strong interactions are incorporated in the hadronic matrix elements
of local four-quark operators. However,
the quark final states are not uniquely related to the physical states,
and the overlap between them could be very complicated.
In order to give reliable predictions of $CP$ asymmetries in
$B$-meson decays, a deeper study of the dynamics of non-leptonic
weak transitions, especially at the hadron level, becomes more urgent today
[19].

	In conclusion, measurements of neutral $B$ decays
at the $\Upsilon (4S)$ supply a valuable opportunity
to probe the sources of $CP$ violation beyond the $K$-meson system and to
advance
our understanding of final-state strong interactions in exclusive weak decays.
In view of recent development in building high-luminosity $B$
factories [4,5], we believe that the work done here should be useful
for experimental studies of $CP$ violation and $B$-meson decays
in the near future. \\

\begin{flushleft}
{\Large\bf Acknowledgements}
\end{flushleft}

	One of us (H.F.) would like to thank Dr. A. Ali for his
useful comments. D.W. acknowledges many interesting conversations
with Drs. D. Judd and D. Wagoner.
Z.X. is indebted to
the Alexander von Humboldt Foundation for its financial support.

\newpage

\begin{figure}
\begin{picture}(400,250)
\put(100,40){\framebox(240,200)}
\put(49,138){${\cal A}_{C}$}
\put(79,136){0}
\put(50,234){$+\sin (2\beta)$}
\put(50,40){$-\sin (2\beta)$}
\put(99,20){$0$}
\put(216,20){$\pi$}
\put(329,20){$2\pi$}
\put(201,0){$\Delta m\Delta t$}
\put(200,140){\line(0,1){15}}
\put(220,140){\line(0,1){15}}
\put(190,147.5){\vector(1,0){10}}
\put(230,147.5){\vector(-1,0){10}}
\put(202,145.5){$\phi_{x_{d}}$}
\put(260,193){\vector(0,1){26}}
\put(260,174){\vector(0,-1){30}}
\put(254,180){$L_{1}$}
\put(295,118){\vector(0,1){20}}
\put(295,102){\vector(0,-1){22}}
\put(290,106){$L_{2}$}
\put(100,215){\line(5,0){5}}
\put(340,215){\line(-5,0){5}}
\put(100,190){\line(5,0){5}}
\put(340,190){\line(-5,0){5}}
\put(100,165){\line(5,0){5}}
\put(340,165){\line(-5,0){5}}
\put(100,115){\line(5,0){5}}
\put(340,115){\line(-5,0){5}}
\put(100,90){\line(5,0){5}}
\put(340,90){\line(-5,0){5}}
\put(100,65){\line(5,0){5}}
\put(340,65){\line(-5,0){5}}
\put(130,40){\line(0,5){5}}
\put(130,240){\line(0,-5){5}}
\put(160,40){\line(0,5){5}}
\put(160,240){\line(0,-5){5}}
\put(190,40){\line(0,5){5}}
\put(190,240){\line(0,-5){5}}
\multiput(220,40)(0,5){40}{\line(0,1){1.5}}
\put(250,40){\line(0,5){5}}
\put(250,240){\line(0,-5){5}}
\put(280,40){\line(0,5){5}}
\put(280,240){\line(0,-5){5}}
\put(310,40){\line(0,5){5}}
\put(310,240){\line(0,-5){5}}
\put(100,140){\circle*{2.5}}
\put(105,127){\circle*{2.5}}
\put(110,114){\circle*{2.5}}
\put(115,102){\circle*{2.5}}
\put(120,90){\circle*{2.5}}
\put(125,79){\circle*{2.5}}
\put(130,69){\circle*{2.5}}
\put(135,61){\circle*{2.5}}
\put(140,54){\circle*{2.5}}
\put(145,48){\circle*{2.5}}
\put(150,44){\circle*{2.5}}
\put(155,41){\circle*{2.5}}
\put(160,40){\circle*{2.5}}
\put(165,41){\circle*{2.5}}
\put(170,44){\circle*{2.5}}
\put(175,48){\circle*{2.5}}
\put(180,54){\circle*{2.5}}
\put(185,61){\circle*{2.5}}
\put(190,69){\circle*{2.5}}
\put(195,79){\circle*{2.5}}
\put(200,90){\circle*{2.5}}
\put(205,102){\circle*{2.5}}
\put(210,114){\circle*{2.5}}
\put(215,127){\circle*{2.5}}
\put(220,140){\circle*{2.5}}
\put(225,153){\circle*{2.5}}
\put(230,166){\circle*{2.5}}
\put(235,178){\circle*{2.5}}
\put(240,190){\circle*{2.5}}
\put(245,201){\circle*{2.5}}
\put(250,211){\circle*{2.5}}
\put(255,219){\circle*{2.5}}
\put(260,226){\circle*{2.5}}
\put(265,232){\circle*{2.5}}
\put(270,236){\circle*{2.5}}
\put(275,239){\circle*{2.5}}
\put(280,240){\circle*{2.5}}
\put(285,239){\circle*{2.5}}
\put(290,236){\circle*{2.5}}
\put(295,232){\circle*{2.5}}
\put(300,226){\circle*{2.5}}
\put(305,219){\circle*{2.5}}
\put(310,211){\circle*{2.5}}
\put(315,201){\circle*{2.5}}
\put(320,190){\circle*{2.5}}
\put(325,178){\circle*{2.5}}
\put(330,166){\circle*{2.5}}
\put(335,153){\circle*{2.5}}
\put(340,140){\circle*{2.5}}
\put(100,94){\circle{3}}
\put(105,85){\circle{3}}
\put(110,78){\circle{3}}
\put(115,71){\circle{3}}
\put(120,66){\circle{3}}
\put(125,61){\circle{3}}
\put(130,59){\circle{3}}
\put(135,58){\circle{3}}
\put(140,57){\circle{3}}
\put(145,59){\circle{3}}
\put(150,62){\circle{3}}
\put(155,66){\circle{3}}
\put(160,71){\circle{3}}
\put(165,78){\circle{3}}
\put(170,85){\circle{3}}
\put(175,94){\circle{3}}
\put(180,103){\circle{3}}
\put(185,113){\circle{3}}
\put(190,124){\circle{3}}
\put(195,135){\circle{3}}
\put(200,146){\circle{3}}
\put(205,156){\circle{3}}
\put(210,167){\circle{3}}
\put(215,177){\circle{3}}
\put(220,186){\circle{3}}
\put(225,195){\circle{3}}
\put(230,202){\circle{3}}
\put(235,209){\circle{3}}
\put(240,214){\circle{3}}
\put(245,219){\circle{3}}
\put(250,221){\circle{3}}
\put(255,222){\circle{3}}
\put(260,223){\circle{3}}
\put(265,221){\circle{3}}
\put(270,218){\circle{3}}
\put(275,214){\circle{3}}
\put(280,209){\circle{3}}
\put(285,202){\circle{3}}
\put(290,195){\circle{3}}
\put(295,186){\circle{3}}
\put(300,177){\circle{3}}
\put(305,166){\circle{3}}
\put(310,156){\circle{3}}
\put(315,145){\circle{3}}
\put(320,134){\circle{3}}
\put(325,124){\circle{3}}
\put(330,113){\circle{3}}
\put(335,103){\circle{3}}
\put(340,94){\circle{3}}
\multiput(100,140)(10,0){25}{\circle*{2.5}}
\multiput(100,76)(10,0){25}{\circle{3}}
\put(175,186){${\cal A}^{\rm A}_{+}(\Delta t)$}
\put(210,104){${\cal A}^{\rm A}_{-}(\Delta t)$}
\put(130,148){${\cal A}^{\rm S}_{-}$}
\put(260,60){${\cal A}^{\rm S}_{+}$}
\end{picture}
\vspace{0.5cm}

	{\bf Fig. 1} $~$ $CP$ asymmetries in $B^{0}_{d}$ versus $\bar{B}^{0}_{d}
\rightarrow J/\psi K_{S}$ at the $\Upsilon (4S)$. Here $x_{d}\equiv \Delta m
/\Gamma$ is a measurable of $B^{0}_{d}$-$\bar{B}^{0}_{d}$ mixing, and
$\phi_{x_{d}}
=\tan^{-1}x_{d}$; $\beta\equiv \arg (-V^{*}_{cb}V_{cd}/V^{*}_{tb}V_{td})$ is an
angle of the CKM unitarity triangle; $L_{1}=|\sin
(2\beta)|/\sqrt{1+x^{2}_{d}}$,
and $L_{2}=2x_{d}|\sin (2\beta)|/(1+x^{2}_{d})^{2}$.

\vspace{4cm}

\begin{picture}(400,300)
\put(70,330){\line(1,0){90}}
\put(70,280){\line(1,0){90}}
\put(160,305){\oval(90,25)[l]}
\put(145,330){\vector(-1,0){2}}
\put(163,325){$\bar{c}$}
\put(145,280){\vector(1,0){2}}
\put(163,275){$d$}
\put(85,330){\vector(-1,0){2}}
\put(62,325){$\bar{b}$}
\put(145,317.5){\vector(1,0){2}}
\put(163,312.5){$u$}
\put(85,280){\vector(1,0){2}}
\put(62,275){$d$}
\put(145,292.5){\vector(-1,0){2}}
\put(163,287.5){$\bar{s}$}
\put(42,300){$B^{0}_{d}$}
\put(172,320){$\bar{D}^{(*)0}$}
\put(172,282){$K^{0} \Longrightarrow K_{S}$}
\multiput(100,330)(3,-5){5}{\line(0,-1){5}}
\multiput(97,330)(3,-5){6}{\line(1,0){3}}
\put(113,260){(a)}
\put(310,330){\line(1,0){90}}
\put(310,280){\line(1,0){90}}
\put(400,305){\oval(90,25)[l]}
\put(325,330){\vector(1,0){2}}
\put(302,325){$b$}
\put(325,280){\vector(-1,0){2}}
\put(302,275){$\bar{d}$}
\put(282,300){$\bar{B}^{0}_{d}$}
\put(385,330){\vector(1,0){2}}
\put(403,325){$u$}
\put(385,280){\vector(-1,0){2}}
\put(403,275){$\bar{d}$}
\put(385,317.5){\vector(-1,0){2}}
\put(403,312.5){$\bar{c}$}
\put(385,292.5){\vector(1,0){2}}
\put(403,287.5){$s$}
\put(412,320){$\bar{D}^{(*)0}$}
\put(412,282){$\bar{K}^{0} \Longrightarrow K_{S}$}
\multiput(340,330)(3,-5){5}{\line(0,-1){5}}
\multiput(337,330)(3,-5){6}{\line(1,0){3}}
\put(353,260){(b)}
\put(70,240){\line(1,0){90}}
\put(70,190){\line(1,0){90}}
\put(160,215){\oval(90,25)[l]}
\put(145,240){\vector(-1,0){2}}
\put(163,235){$\bar{u}$}
\put(145,190){\vector(1,0){2}}
\put(163,185){$d$}
\put(85,240){\vector(-1,0){2}}
\put(62,235){$\bar{b}$}
\put(145,227.5){\vector(1,0){2}}
\put(163,222.5){$c$}
\put(85,190){\vector(1,0){2}}
\put(62,185){$d$}
\put(145,202.5){\vector(-1,0){2}}
\put(163,197.5){$\bar{s}$}
\put(42,210){$B^{0}_{d}$}
\put(172,230){$D^{(*)0}$}
\put(172,192){$K^{0} \Longrightarrow K_{S}$}
\multiput(100,240)(3,-5){5}{\line(0,-1){5}}
\multiput(97,240)(3,-5){6}{\line(1,0){3}}
\put(113,170){(${\rm a}^{'}$)}
\put(310,240){\line(1,0){90}}
\put(310,190){\line(1,0){90}}
\put(400,215){\oval(90,25)[l]}
\put(325,240){\vector(1,0){2}}
\put(302,235){$b$}
\put(325,190){\vector(-1,0){2}}
\put(302,185){$\bar{d}$}
\put(282,210){$\bar{B}^{0}_{d}$}
\put(385,240){\vector(1,0){2}}
\put(403,235){$c$}
\put(385,190){\vector(-1,0){2}}
\put(403,185){$\bar{d}$}
\put(385,227.5){\vector(-1,0){2}}
\put(403,222.5){$\bar{u}$}
\put(385,202.5){\vector(1,0){2}}
\put(403,197.5){$s$}
\put(412,230){$D^{(*)0}$}
\put(412,192){$\bar{K}^{0} \Longrightarrow K_{S}$}
\multiput(340,240)(3,-5){5}{\line(0,-1){5}}
\multiput(337,240)(3,-5){6}{\line(1,0){3}}
\put(353,170){(${\rm b}^{'}$)}
\put(0,130){{\bf Fig. 2} $~$ Quark diagrams for $B^{0}_{d}$ versus
$\bar{B}^{0}_{d}$
decays to $\bar{D}^{(*)0}K_{S}$ and $D^{(*)0}K_{S}$.}

\end{picture}
\end{figure}

\end{document}